\documentclass[letter,oldversion]{aa}

\usepackage{graphicx}
\usepackage{epsfig}
\usepackage{color}
\usepackage{txfonts}
\usepackage{natbib}
\bibpunct{(}{)}{;}{a}{}{,} 

\begin{document}

\title{Generation of rotationally dominated galaxies by mergers of pressure-supported progenitors}
\titlerunning{Rotationally dominated galaxies by mergers of pressure-supported progenitors}
\author{Paola Di Matteo$^{1}$, Chanda J. Jog$^{2}$, Matthew D. Lehnert$^{1}$, Fran\c coise Combes$^{3}$, Benoit Semelin$^{3}$
}
\authorrunning{Di Matteo et al.}
\institute{
$^{1}$  Observatoire de Paris, section de Meudon, GEPI,  5 Place Jules Jannsen, 92195, Meudon, France\\
$^2$ Department of Physics, Indian Institute of Science, Bangalore 560012, India\\
$^3$ Observatoire de Paris, LERMA, 61 Avenue de l'Observatoire, 75014 Paris, France
}

\date{Accepted,
      Received }


\abstract{Through the analysis of a set of numerical simulations
of major mergers between initially non-rotating, pressure supported
progenitor galaxies with a range of central mass concentrations, we
have shown that: (1) it is possible to generate elliptical-like galaxies, with $v/\sigma > 1$
outside one effective radius, as a result of the conversion of orbital-
into internal-angular momentum;
(2) the outer regions acquire part of the angular momentum
first; (3) both the baryonic and the dark matter components of the remnant
galaxy acquire part of the angular momentum, the relative fractions depend on the
initial concentration of the merging galaxies. 
For this conversion to occur the initial baryonic
component must be sufficiently dense and/or the encounter should
take place on a orbit with high angular momentum.  Systems with these
hybrid properties have been recently observed through a combination of
stellar absorption lines and planetary nebulae for kinematic studies of
early-type galaxies. Our results are in qualitative agreement with such
observations and demonstrate that even mergers composed of non-rotating,
pressure-supported progenitor galaxies can produce early-type galaxies
with significant rotation at large radii.}


\maketitle

\section{Introduction}\label{intro}

It has been known for several decades that early type galaxies
have complex kinematics and varying amounts of rotation and that
these characteristics depend on luminosity and isophotal shape
\citep[e.g.,][]{davies83, bender88}.  Recently, the results of the
SAURON survey \citep{bacon01, dezeeuw02} have lead to a division of
early-type galaxies into two distinct classes, slow and fast rotators,
depending on the amount of the angular momentum per unit mass of
their stellar component inside one effective radius \citep[$R_e$;
][]{emsellem07}. These two classes also have different properties, slow
rotators are generally more massive, less flattened systems than fast
rotators \citep{cappellari07}. 
Studies of the more distant regions of
early-type galaxies, traced by planetary nebulae, have shown that their complexity of the kinematics
of the inner regions also extends to the outer halos \citep{coccato08}.
Interestingly, these outer haloes can be more rotationally dominated
than their central regions \citep{coccato08}.

Extensive studies using numerical simulations suggest that different
processes lead to the formation of slow and fast rotating early-type
galaxies. Most of these studies have focused on the remnants of
spiral-spiral mergers, which is likely to be an important process for the
evolution of formation of ellipticals and perhaps responsible for driving
galaxies from the blue cloud to the red sequence \citep{cattaneo06,
faber07, romeo08}.  Even if the agreement between simulations and
observations is often qualitative only - which is most likely due to
incomplete knowledge of the characteristics of the progenitors of these
merger remnants - some conclusions can still be drawn \citep{quinn93,
naab99, velaz99, bendo00, cretton01, naab03, bournaud04, bournaud05,
naab06a, gonzalez05a, gonzalez06, jesseit07, jesseit08}. 
Mergers of two spirals can produce
rotationally supported early-type galaxies (at radii inside 1 $R_e$),
depending on the mass ratio of the progenitors.  Major mergers
(1:1 mass ratio) are sufficiently violent to completely destroy the
initial ordered rotational motions in the progenitor disks\footnote{Unless
peculiar configurations are considered \citep{pfenniger97, puerari01}
or the amount of gas in the progenitor disks is sufficiently high
\citep{springel05, robertson06, hopkins09}.}, typically producing a
pressure-supported system.  Apparently, mergers with higher mass ratios
can generally produce remnants with significant rotation.  Virtually all
the focus so far has been on simulated
spiral-spiral interactions, attempting to answer to the question: Is it
possible to (at least partially) preserve the initial internal angular
momentum of the progenitor disks?

Encounters between pressure supported (spheroid dominated) galaxies
have been systematically studied by \citet{white78, white79, gonzalez05b,
gonzalez05c, gonzalez06}. These studies show that such mergers
result in a wide variety of remnant
properties.  In particular, \citet{gonzalez05b} pointed out that mergers
of spherically symmetric galaxies, without dark matter, can result in
remnants partially supported by rotation and  that part of the orbital angular momentum is absorbed by the halo, when present \citep{gonzalez05c}. However, none of these
studies have investigated in detail how the orbital angular momentum is
redistributed during the interaction, nor have they investigated the
dynamical properties of the remnants out to large distances (several
$R_e$).

In this Letter, we show, for the first time, that mergers of initially
spherical, pressure-supported stellar systems without rotation can produce
hybrid remnants having elliptical-like morphology, but are
dominated by rotation ($v/\sigma > 1$). As we shall demonstrate,
the final characteristics of the remnant depends on both the initial
orbital angular momentum and the central density of the merging galaxies.

\section{Models and initial conditions}\label{init}

We study the coalescence of two equal-mass elliptical galaxies,
consisting of a stellar and a dark matter (DM) component, distributed in
a Plummer density profile, without any dissipational component and initial rotation.  The density profile
of the dark matter halo is the same in all the simulations, we only
changed the central density of the baryonic component in order to study
how changing the density affects the angular momentum (AM) redistribution
during the encounter. We thus considered gE0l, gE0 and gE0m models, with increasing central stellar density (Table~\ref{galpar} and Fig.~\ref{initfig}), whose half-mass radii $r_{50}$ and effective densities $\sigma^2/{r_{50}}^2$ are in agreement with observational estimates for galaxies with $M_r=-22$ \citep{desro07}, assuming a M/L=3 in the r-band . Since the total mass of the stellar component is fixed, increasing the central stellar density results in a more concentrated and compact stellar profile.  In these models, the DM contributes $\sim$ 10-30$\%$ of the total mass within the half-mass radius of the baryonic component, in agreement with observational estimates \citep{barnabe09}. The initial galaxies have been then placed at
a relative distance of 100 kpc, with a variety of relative velocities,
in order to simulate different orbits (orbital energies and angular
momenta; see \citet{dimatteo09}).  In particular, in the following we will show results from three of these orbits (id=01, 05, 15), whose main parameters are given in Table~\ref{orbite}. When refering to specific encounters, the nomenclature adopted is the following: morphological type of the two galaxies in the interaction (gE0, gE0l or gE0m), plus the encounter identification string (see first column in Table 2). Each pair has been modeled with $N=120000$
particles, distributed among stars ($N_{star}=80000$) and dark matter
($N_{DM}$=40000).

 All the simulations have been run using the Tree-SPH code described
in \citet{benoit02}.  A Plummer potential is used to soften the gravity at small
scales, with constant softening lengths  of $ \epsilon=280\ \mathrm{pc}$
for all particles. The equations of motion are integrated using a leapfrog
algorithm with a fixed time step of 0.5~Myr. With these choices, the relative error in the conservation of the total energy is of the order of  $2\times 10^{-7}$ per time step.
Since the work presented here
only investigates dry-mergers, only the part of the code evaluating
the gravitational forces acting on the systems has been used.

\section{Results}\label{results}
\subsection{Angular momentum redistribution: dependence on the orbits}\label{orbits}

\begin{figure}[h]
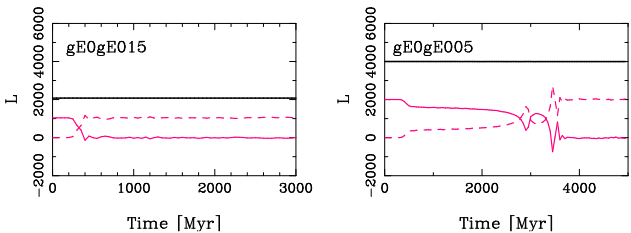

 \begin{minipage}{0.23\textwidth}
   \centering
\includegraphics[width=3.cm,angle=270]{fig3.ps}
 \end{minipage} 
 \begin{minipage}{0.23\textwidth}
   \centering
\includegraphics[width=3.cm,angle=270]{fig4.ps}
 \end{minipage} 
\caption{Evolution of the total (black line), orbital (solid red line) and
internal (dashed red line) AM for two different orbits. The
two galaxies are identical, so the internal and orbital AM are shown only
for one of the two. The angular momenta are all measured perpendicular to the orbital plane and are in units of $2.3\times 10^{11}
 \mathrm{M_{\odot}}$ kpc \mbox{km} \mbox{s}$^{-1}$. \label{angorbit}}
\end{figure}

The  first question we would like to address is: how
 orbital AM is redistributed and converted into
internal AM during an interaction and how does this depend
on the orbital parameters of the encounter?  In Fig.~\ref{angorbit}, we
show the evolution of the total, orbital and internal AM,
for two different orbits (id=15 and id=05), corresponding, respectively,
to increasing initial orbital angular momenta.  The AM is evaluated with respect to the barycenters of the two galaxies, considering all the particles. The main results are:

\begin{figure}[h]
 \begin{minipage}{0.2\textwidth}
   \centering
\includegraphics[width=3.cm,angle=0]{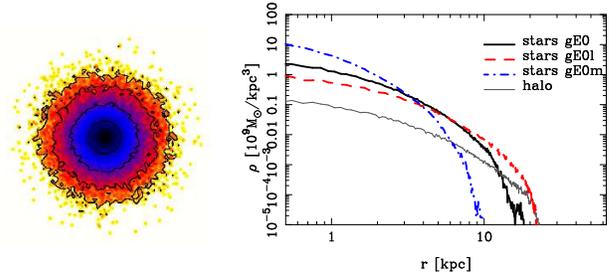}
 \end{minipage} \hspace{-0.2cm}
 \begin{minipage}{0.1\textwidth}
   \centering
\includegraphics[width=3.6cm,angle=270]{fig2.ps}
\end{minipage} \hspace{3.6cm}

\caption{Initial properties of the galaxy models: (a) projected density
map of the stellar component of the gE0 galaxy; (b) volume-density
profiles of the galaxy models. The model is as indicated in the legend of the
figure.
 \label{initfig}}
\end{figure}

\begin{table}[h]
\caption[]{Parameters for the standard (gE0), less concentrated
(gE0l) and  more concentrated (gE0m) model galaxies. Both the stellar and dark
matter profiles are represented by Plummer models, having characteristic
masses, respectively, given by $M_{B}$ and $M_{H}$ for the baryons and halo
respectively, and core radii given
by $r_{B}$ and $r_{H}$ for the baryons and halo respectively.  The initial half-mass radii $r_{50}$ of the baryonic component and effective densities $\sigma^2/{r_{50}}^2$ are also given.}
\label{galpar}
 \centering
\begin{tabular}{lcccc}
 \hline\hline
& $M_{B}, M_{H}$ &  $r_{B}, r_{H}$ & $r_{50}$& $log(\sigma^2/{r_{50}}^2)$  \\
&  $[2.3\times 10^9 M_{\odot}]$  & [kpc] & [kpc]& [(km/s/kpc)$^2$] \\
 \hline 
gE0 & 70, 30 & 4.0, 7.0& 5.2 & 3.2\\
gE0l& 70, 30& 6.0, 7.0& 7.5& 2.9\\
gE0m& 70, 30& 2.0, 7.0& 2.8& 3.9\\
       \hline  
       \hline
  \end{tabular}
\end{table}

   \begin{table}[h]
     \caption[]{Orbital parameters}
     \label{orbite}
     \centering
     \begin{tabular}{ccccc}
       \hline\hline
       orbit id & $r_{ini}^{\mathrm{a}}$ &  $v_{ini}^{\mathrm{b}}$  & $L^{\mathrm{c}}$  & $E^{\mathrm{d}}$ \\
         & [kpc] & [100 $kms^{-1}$] &  [$10^2kms^{-1}kpc$] &  [$10^4km^2s^{-2}$]\\
       \hline
       01& 100.& 2.0&57.0&0.\\
       05& 100.& 2.0  &80.0$^{\mathrm{e}}$&0.\\
       15& 70.0& 1.6&41.3&-1.57\\       
       \hline  
       \hline
     \end{tabular}
\begin{list}{}{}
\item[$^{\mathrm{a}}$] The initial distance between the two galaxies.
\item[$^{\mathrm{b}}$] The absolute value of the initial relative velocity
between the two galaxies.
\item[$^{\mathrm{c}}$] The absolute value of the orbital angular momentum
of the unit mass, i.e., $L=\mid\bf{r_{ini}} \times \bf{v_{ini}}\mid $.
\item[$^{\mathrm{d}}$] The total energy of the relative motion,
i.e., \\ $E={v_{ini}}^2/2-G(m_1+m_2)/r_{ini}$, with $m_1=m_2=2.3
\times10^{11}M_{\odot}$.
\item[$^{\mathrm{e}}$] Even if the initial relative velocities in orbit
05 and 01 are equal, their orbital angular momenta differ because the
initial orbital tangential velocities are different.
\end{list}
\end{table}

\begin{itemize}
\item The total AM is (obviously) conserved during the
interaction.
\item The orbital AM is constant until the first pericenter
passage between the two galaxies. When this happens, dynamical
friction and tidal torques act efficiently  on the systems, converting
part of the orbital AM into internal  AM.
\item At every successive close passage between the two galaxies, part
of the orbital AM is converted into internal rotation of
the two systems. This process ends when all the orbital AM
is converted into internal AM and the two galaxies finally merge.
\item Instead of defining the merging time as the point when the two
galaxy centres are sufficiently close  we suggest, based on dynamical
arguments, that the merging time could be the point when orbital AM
is totally converted into internal AM (as we verified, these two definitions give similar merging times). This definition, of course,
applies only for orbits having an initial orbital AM different from zero.
\item Because of AM conservation, the final amount of internal AM of the
remnant galaxy depends on the initial amount of orbital AM.
\end{itemize}

\begin{figure}[h]
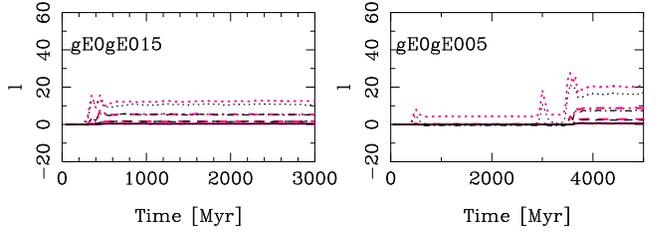

 \begin{minipage}{0.23\textwidth} 
   \centering
\includegraphics[width=3.cm,angle=270]{fig5.ps}
 \end{minipage} 
 \begin{minipage}{0.23\textwidth}
   \centering
\includegraphics[width=3.cm,angle=270]{fig6.ps}
 \end{minipage} 
\caption{Evolution of the specific AM, $l$, for four
different regions of one of the two elliptical galaxies ($r\le 2 kpc$,
solid lines, $2 kpc < r \le 5 kpc$ dashed lines, $5 kpc < r \le 10 kpc$,
dot-dashed lines, and $10 kpc < r \le 20 kpc$, dotted lines. The specific
AM of the stellar component is shown in red, that of the
dark matter component in gray. $l$ is in units of 100 kpc \mbox{km} \mbox{s}$^{-1}$. In
the two panels, the initial orbital parameters of the interaction has
been varied,  so to have an increasing orbital AM,  going from left to
right, while the morphology of the interacting galaxies has been kept
fixed. The specific angular momenta are all measured perpendicular to the orbital plane.\label{angzone}} \end{figure}

In Fig.~\ref{angzone}, we show the evolution with time of the specific
internal AM for both the stellar and the dark matter components.
For each of the two galaxies, and for each component (baryons and halo),
we evaluate the specific internal AM in four different radial regions. From this analysis, we can deduce that:

\begin{itemize}
\item  \emph{The orbital AM is converted into internal AM
``outside-in'', the external regions acquire part of the angular
momentum first}.

\item Interestingly, between the first pericenter passage and the final
phases of coalescence, the outer regions show some rotation, while the
inner regions do not.

\item Both the baryonic and the dark matter component acquire part of
the orbital AM, and both the components, in the final remnant, show
some rotation. \emph{The two initially non-rotating and spherically
symmetric components are thus transformed into a flattened rotating
system\footnote{As we confirmed, the remnants show some
velocity anisotropy and both rotation and anisotropy contribute to the
flattening. A discussion on this will be presented in a more detailed
paper.}}.

\item The higher the initial orbital AM, the higher the
internal specific AM of the final remnant, at all radii.

\end{itemize}

\begin{figure}[h]
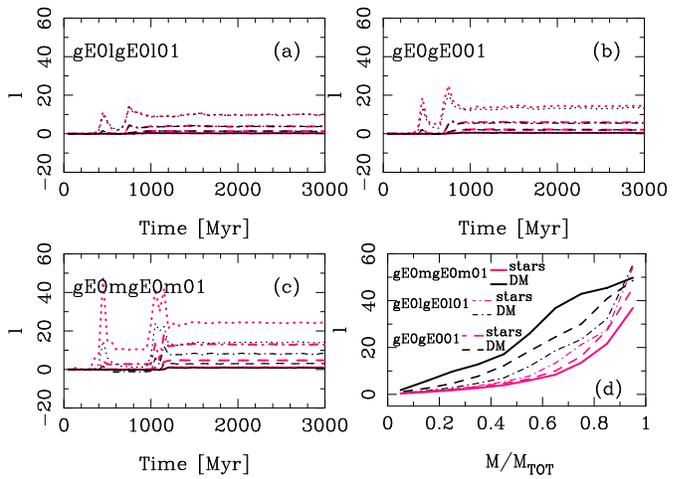

 \begin{minipage}{0.23\textwidth}
   \centering
\includegraphics[width=3.1cm,angle=270]{fig7.ps}
 \end{minipage} 
 \begin{minipage}{0.23\textwidth}
   \centering
 \hspace{-0.8cm}
\includegraphics[width=3.1cm,angle=270]{fig8.ps}
 \end{minipage} 
 \begin{minipage}{0.23\textwidth}
   \centering
\includegraphics[width=3.1cm,angle=270]{fig9.ps}
 \end{minipage} 
 \begin{minipage}{0.23\textwidth}
   \centering
\includegraphics[width=3.1cm,angle=270]{fig10.ps}
 \end{minipage} 
\caption{Panels (a), (b) and (c) show results simular to Fig.~\ref{angzone}, but now
the panels show the evolution of the specific AM
for interacting pairs having initially the same orbital parameters,
but different morphologies (the stellar central
densities increase from panel a-c). Panel (d): Specific
AM, $l$, of the remnant galaxies as a function of radii containing a fixed percentage of the baryonic (red lines) and halo (black lines) mass. $l$
has been evaluated at least 1 Gyr after the coalescence of the two
progenitors. Line-styles correspond to the different
morphologies shown in panels (a), (b) and (c), i.e., gE0lgE0l01
(dot-dashed line), gE0gE001 (dashed line), gE0mgE0m01
(solid line).  \label{angdensity}}
\end{figure}

\subsection{Angular momentum redistribution: dependence on the galaxy
central density}\label{density}

To investigate how the evolution and redistribution of AM depend on
the initial central density of the two progenitors, we also simulated
mergers with a range of concentrations (gE0l-gE0 - less concentrated and
gE0m-gE0m - more concentrated).  Fig.~\ref{angdensity} shows the evolution
with time of the specific internal AM for four different regions in the
galaxy (see also Fig.~\ref{angzone}).  Since the initial orbital AM are
the same in these simulations, the amount of internal AM of the final
remnant will also be the same. However, AM is redistributed in a quite
different way, depending on the initial concentration
of the progenitor galaxies. We find that mergers with higher initial galaxy
central density produce remnants with larger relative amount of rotation
at any radius within a radius of 20 kpc. This result is true for both
the baryonic and dark matter components.

This behavior is related to the fact that the three remnant galaxies
have different mass distribution profiles, as we will discuss in
the next Section, and so ultimately the physical regions chosen in
Fig.~\ref{angdensity} contain different amount of baryonic and dark
matter mass.  A more appropriate comparison is the distribution of
the specific AM within a constant fraction of the mass.
We show the distribution
of the specific AM as a function of radii containing
a fixed percentage of the baryonic (red lines) or dark matter (black
lines) mass in Fig.~\ref{angdensity}, panel (d). Again, this is evaluated at least 1 Gyr after the merger has
completed. The internal rotation is distributed in
a quite different way in the three remnants as a function of mass. The
baryons in each of the models have a similar distribution of $l$ inside the
effective radius, while outside it, less concentrated progenitors lead
to a higher amount of specific AM. The dark matter component, however,
shows the opposite trend -- the specific AM is higher for all values
of the enclosed mass for progenitors with high concentration.  This is
because more concentrated baryonic components are less  responsive to
tidal torques, while the outer, less concentrated halo is more strongly
affected by the tidal interaction. To illustrate this, in the case of
the gE0lgE0l01 interaction, at the end of the simulation about
$67\%$ of the orbital AM has been acquired by the baryons,
and only $33\%$ by the dark halo, while for the most concentrated case
(gE0mgE0m01) the two components have acquired similar percentage of
the initial orbital AM ($48\%$ for the baryons and $52\%$ for the halo).\\
Thus \emph{baryonic components with low concentration are more susceptible to
tidal torques, and thus acquire a higher percentage of the initial
orbital angular momentum}.

\subsection{Mergers of dense pressure-supported systems: rotationally-dominated systems with an elliptical-like
morphology}\label{hybrid}

\begin{figure}
\begin{minipage}{0.2\textwidth}
\centering
\vspace{-0.6cm}
\includegraphics[width=2.7cm,angle=0]{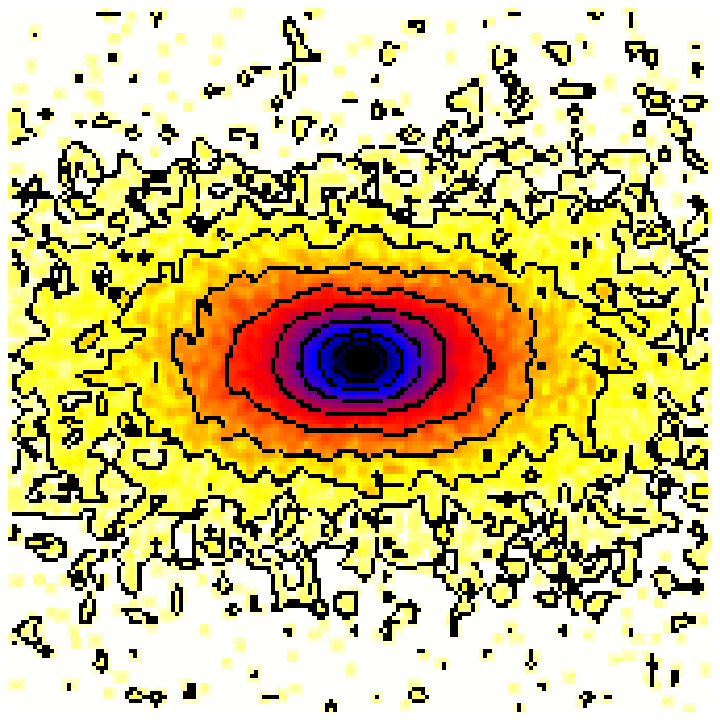}
\end{minipage} 
\begin{minipage}{0.2\textwidth}
\hspace{-1.6cm}
\includegraphics[width=3.2cm,angle=270]{fig12.ps}
\end{minipage} 
\begin{minipage}{0.2\textwidth}
\centering
\vspace{-0.6cm}
\includegraphics[width=2.7cm,angle=0]{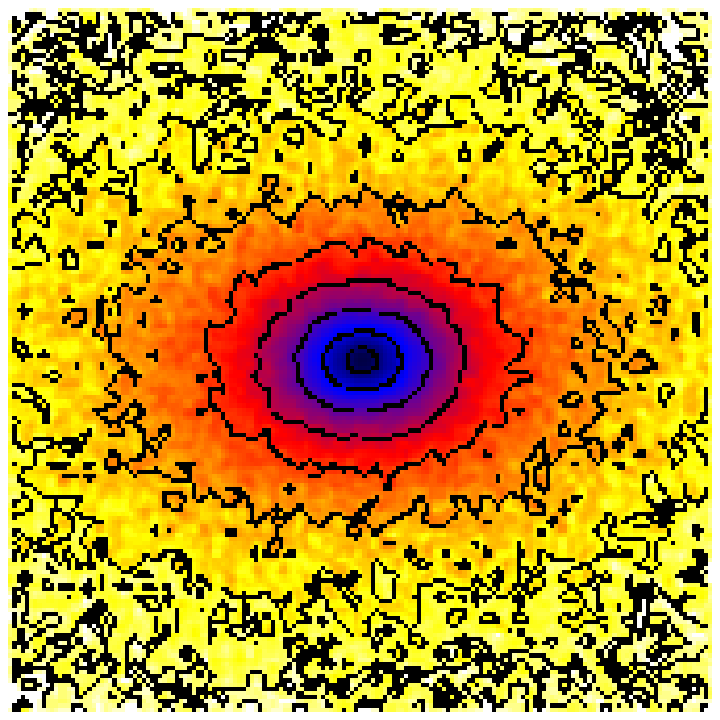}
\end{minipage} 
\begin{minipage}{0.2\textwidth}
\includegraphics[width=3.2cm,angle=270]{fig14.ps}
\end{minipage} 
\caption{Morphological characteristics of two remnants of
pressure-supported, initially non-rotating galaxies. Top panels:  (Left)
Projected density map of the remnant of the gE0mgE0m01 encounter
(progenitors with high concentrations); (right) Surface density profile
of the baryonic (red color) and dark matter component (black color) of
the remnant (solid line) and its progenitor galaxy (dashed line). The
green arrows indicate the half mass radius ($r_{50}$) of the baryonic
component. Bottom panels: Same as top panels, but progenitors with
relatively low concentrations (gE0lgE0l01).
\label{morph}}
\end{figure}

\begin{figure}
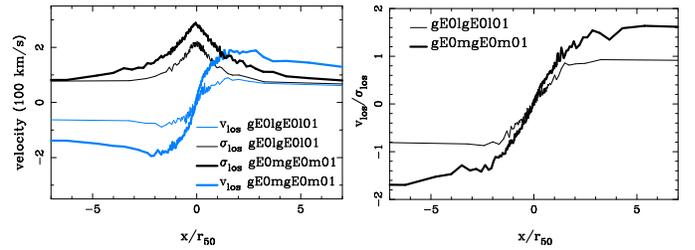

\includegraphics[width=3.2cm,angle=270]{fig15.ps}
\includegraphics[width=3.2cm,angle=270]{fig16.ps}
\caption{ Kinematical properties of two remnants of E-E mergers.
Left panel: line-of-sight velocities (blue) and velocity dispersions (black) of the gE0mgE0m01 (thick lines) and of the gE0lgE0l01 (thin lines) remnants. Right panel: corresponding $v/\sigma$ ratio (gE0mgE0m01, thick line;  gE0lgE0l01; thin line). \label{vlos2}}
\end{figure}

The varying amount of specific AM found in remnants of progenitors having
different initial concentrations is also reflected in the kinematics.
A detailed analysis of the relationship between the morphological and
dynamical properties of mergers of pressure-supported systems will be
the topic of a future paper. Here we simply note that when a sufficient
amount of specific AM is imparted to the central regions of the remnant
galaxy, such remnants may show interesting hybrid properties. Some may
have a morphology typical of an elliptical galaxy, but a $v/\sigma$
ratio higher than 1, over most of their extended stellar
distribution. This is the case, for example, in the gE0mgE0m01 merger
remnant.  Even if its baryonic component has a specific AM inside the
half-mass radius similar to that of the less concentrated galaxies
(panel d of Fig.~\ref{angdensity}), the gE0mgE0m01 remnant is much
denser than, for example, the gE0lgE0l01 remnant 
(Fig.~\ref{morph}). The different mass distribution of the two galaxies
is reflected in different dynamical properties in an interesting
way:  even if they acquire a lower percentage of the orbital AM,
being less susceptible to tidal torques, the baryonic components of
the most concentrated systems show a higher central velocity dispersion,
a steeper dispersion profile, a higher value of the line-of-sight
velocities, and a $v/\sigma$ ratio which becomes greater than 1 at
about $r=r_{50}$ (the baryonic half-mass radius; Fig.~\ref{vlos2}).
This may appear paradoxical, but this is due to the fact that
the internal AM is redistributed over a much more compact distribution
compared to low-concentration remnants.  Most of the galaxies studied
by \citet{coccato08} show  a $v/\sigma$ ratio not greater than
0.6, while a fraction of them have kinematics that become increasingly
supported by rotation in the outer parts, qualitatively in agreement
with the hybrid merger remnants discussed here. This supports the idea that
the morphological properties of a galaxy are not univocally related to
the dynamical ones. Mergers can indeed produce hybrid systems, having a
spiral-like morphology but an elliptical-like kinematics  \citep{jog02,
bournaud04, bournaud05} or, as presented here, elliptical-like morphology
but are rotationally-dominated.


\section{Conclusions}\label{conclusions}

Through the analysis of a set of numerical simulations of major mergers
between initially non-rotating, pressure supported progenitor galaxies,
we have shown that it is possible to generate elliptical-like galaxies, with $v/\sigma
> 1$ outside one effective radius, simply as a result of the  conversion
of the orbital AM into internal one, during the mergers of two initially
non-rotating progenitors. For this to occur the initial baryonic component
must be sufficiently dense and/or the encounter should take place on a
high AM orbit.

Systems with such hybrid properties have been recently observed.
For example, \citet{coccato08} found that galaxies in their sample with
the most steeply declining velocity profiles have generally fast rotating
halos, as found for our fast rotating remnants. Our results are in qualitative agreement with such
observations and demonstrate that even mergers composed of non-rotating,
pressure-supported progenitor galaxies can produce early-type galaxies
with significant rotation at large radii. In particular, we stress the important point that this mechanism does not
require initially any internal AM in the colliding galaxies.

Simulations by \citet{naabkho06} show that repeated binary
early-type mergers lead to the formation of anisotropic, slowly rotating
elliptical galaxies (inside $R_e$). In this context, whether the hybrid
systems presented in this Letter can be formed frequently and maintained
also after successive mergers should be studied with a larger set of
simulations, taking into account repeated mergers (major and minor) with
different relative inclinations, as well as progenitors with different
DM profiles \citep{nav97}.

Nevertheless, it is possible to hypothesize a scenario where these
elliptical, rotationally-supported systems are mainly found among remnants
of dense early-type galaxies. These hybrid features will tend to vanish
if successive major mergers take place, with a random orientation of
the internal and orbital AM with respect to the spin of the "hybrid"
galaxy. But this has yet to be investigated numerically.

\section*{Acknowledgments} 
PDM  was awarded a travel grant from the Indo-French Astronomy Network which made possible a visit to IISc, Bangalore, in November 2008. PDM is supported by a grant from the Agence Nationale de la Recherche (ANR) in France. We thank the referee for comments which helped to improve the paper.

\end{document}